\documentstyle[prl,aps,manuscript]{revtex}
\begin{document}
\draft

\title{EFFECT OF FLOW ON CALORIC CURVE FOR FINITE NUCLEI }

\author{S. K. Samaddar$^{1)}$, J. N. De$^{2)}$, 
     and S. Shlomo$^{3)}$}

\address{$^{1)}$ Saha Institute of Nuclear Physics, 1/AF,
Bidhannagar, Calcutta - 700064, India}
\address{$^{2)}$ Variable Energy Cyclotron Centre, 1/AF,
Bidhannagar, Calcutta - 700064, India}
\address{$^{3)}$ Cyclotron Institute, Texas A\&M University,
College Station, TX 77843-3366, USA}

\maketitle

\begin{abstract}
In a finite temperature Thomas-Fermi theory, we construct caloric 
curves for finite nuclei enclosed in a freeze-out volume few times 
the normal nuclear volume, with and without inclusion of flow. 
Without flow, the caloric curve indicates a smooth liquid-gas phase 
transition whereas with flow, the transition may be very sharp. We 
discuss these results in the context of two recent experiments, one 
for heavy symmetric system (Au + Au at 600A MeV) and the other for 
highly asymmetric system (Au + C at 1A GeV) where different
behaviours in the caloric curves are seen. 
\end{abstract}

\vskip 1 cm

\vskip 1 cm


\newpage

Liquid-gas phase transition in nuclei is one of the fundamental issues 
driving intense studies in medium and high energy nuclear collisions 
in recent times. Theoretical studies on the equation of state (EOS) for 
infinite nuclear matter showed criticality at temperatures 
$ T_c \sim 15-20 $ MeV \cite {KWH,JMZ,BSSD} depending on the 
interaction chosen and possible signals of such a critical
behaviour were surmised from the power law 
distribution \cite{Finn,GKJ,Chit,Lyn}
of mass or charge fragments detected in energetic nuclear collisions. A
clearer signal of liquid-gas phase transition in nuclei is hinted at from
the experimental caloric curve \cite{Poch} obtained in Au + Au collisions 
at 600A MeV where in the excitation energy range of $\sim$ 4-10 MeV per
particle, the temperature $T$ is found to be almost constant at a value of 
$T \sim 5$ MeV, even if there may be some uncertainty in the extraction of
temperatures \cite{CKP}. The specific heat at constant volume would then 
show a very sharp peak at this temperature and the excitation energy range 
over which $T$ remains constant could be termed as the latent heat of 
vaporisation. On the theoretical side, fragmentation calculations in the 
microcanonical algorithm of Gross \cite{Gro} and in the Copenhagen 
canonical description \cite{BDMS,BBI} show such a structure in the 
specific heat signalling a liquid-gas phase transition. The lattice-gas
model for fragmentation \cite{DPKG} also shows such a peak. These trends 
in the caloric curve were seen in a very recent calculation \cite{DDSS} 
in the finite temperature Thomas-Fermi (FTTF) theory. Contrary to the 
extremely sharp peak at $T \sim 5$ MeV as inferred from the experimental 
caloric curve \cite{Poch}, the FTTF calculation showed a broad-based 
peak at $T\sim 9-10$ MeV depending on the system chosen.

Another set of experimental data with Au on C is reported recently 
\cite{Hau} at a little higher energy (1A GeV) where the caloric curve 
is somewhat different; here around 6-7 MeV the temperature rises slowly 
but steadily with excitation energy, reminiscent of a continuous phase 
transition. These two different behaviours in the caloric curve reflect 
subtle changes in the physical process at different bombarding energies 
and for different systems and in the following, we try to explore and 
understand it within the framework of FTTF theory \cite{DRPS}.

For a description of the hot nuclear material formed in energetic 
nuclear collisions, we consider a nucleus of $N_0$ neutrons and $Z_0$ 
protons, in thermodynamic equilibrium at a temperature $T$ within a 
'freeze-out' volume $V$. In a generalised description, the systems may 
be initially compressed and because of decompression may have collective 
radial flow in addition to thermal excitation. An expanding system, in 
a strict thermodynamic sense, is not in equilibrium. However, if the 
time scale involved in the expansion is much larger compared to the 
equilibration times in the expanding complex, {\it i.e.} the flow 
velocity is quite small compared to the average nucleonic velocity, the 
assumption of thermodynamic equilibrium may not be inappropriate. In a 
recent paper, Pal et al \cite{PSD} suggested to simulate the effect of 
collective radial flow  through inclusion of an external negative 
pressure in the total thermodynamic potential at freeze-out volume. In 
absence of flow, at the freeze-out, the kinetic contribution of the 
thermal pressure is generally assumed to be cancelled by the interaction 
contribution, i.e, the system is at equilibrium under zero external 
pressure. A positive uniform external pressure gives rise to compression; 
similarly a negative external pressure gives rise to decompression 
resulting in the outward radial flow of matter. The expanding system can 
hence be assumed to be under the action of a negative external pressure 
$P_0$ , whose magnitude is equal to the flow pressure $P_f (|P_0|= P_f)$, 
the internal pressure exerted by the radially outgoing nucleons at the 
freeze-out surface. The total thermodynamic potential of the system 
\cite{DRPS} at freeze-out is given by 
$$
G = E - TS - \mu_n N_0 - \mu_p Z_0 + P_0\Omega, 
\eqno (1)
$$
where E and S are the energy and entropy of the system, respectively,
$\mu_n$ and $\mu_p$ are the chemical potentials of neutrons and protons,
respectively, $ P_0 $ the constant external
pressure assumed negative and $\Omega$ the effective volume given by
$$
\Omega = \frac{4}{3}\pi R_u^{3},\,\,\,\,\,\,\, 
R_u = (\frac{5}{3}<r^2>)^{1/2}.  
\eqno (2)
$$
Here $<r^2>^{1/2}$ is the root mean square radius (rms) of the matter
density distribution and $R_u$ is the radius of the corresponding 
uniform density distribution. We have taken this value of $\Omega$
in the present calculations. The interaction density is calculated 
with a Seyler-Blanchard type momentum and density dependent finite 
range two-body effective interaction \cite{DRPS}; the Coulomb part 
with direct and exchange
terms are included in this interaction density. Minimisation of the
thermodynamic potential in the Thomas-Fermi approximation then leads 
to the expression for occupation probability as
$$
n_\tau(r,p) = [1+exp\{(\frac{p^2}{2m_\tau^\ast(r)} + V_\tau^0 (r) +
V_\tau^2(r) -\mu_\tau + P_0 \frac{10 \pi }{3 A} R_u r^2)/T\}]^{-1}.  
\eqno (3)
$$
Here $\tau$ refers to neutrons or protons, 
$V_\tau^0$ is the single 
particle potential (which includes the Coulomb term for protons), 
$m_\tau^\ast$ the effective mass and $V_\tau^2$ the rearrangement 
potential that appears for a density dependent interaction. The density 
at any point is obtained from the momentum integration of the occupancy. 
The total energy density at temperature $T$ is then written as 
$$
\varepsilon(r)=\sum_{\tau} \, \rho_{\tau}(r) [T \, J_{3/2}(\eta_{\tau}(r))/
J_{1/2}(\eta_{\tau}(r)) {(1- m_{\tau}^{\ast}(r) V_{\tau}^{1}(r))}
+\frac{1}{2}V_{\tau}^{0}(r)]. 
\eqno (4)
$$
In Eq. (4), $J's$ are the usual Fermi integrals, $V_\tau^1$ is the 
potential term that comes with momentum dependence and is associated with 
$m_\tau^\ast$. The fugacity $\eta_\tau(r)$ is defined as
$$
 \eta_{\tau}(r)=[\mu_{\tau}-V_{\tau}^{0}(r)-V_{\tau}^{2}(r) -
P_0\frac{10 \pi }{3 A} R_u r^2]/T. 
\eqno (5)
$$

Once the interaction energy density is known, the nuclear density can be
obtained self-consistently. The total energy per particle is then given by
$$
e(T) = \int\varepsilon (r) \,d{\bf r}/A. 
\eqno (6)
$$
For details on the FTTF theory, we refer to Ref.\cite{DRPS}.

For a finite system at nonzero temperature, the continuum states are
occupied with a finite probability as a result of which the particle 
density does not vanish at large distances. The observables then depend 
on the size of the box in which the calculations are done. Guided by the 
practice that many calculations for nuclear collisions are done by 
imposing that thermalisation occurs in a freeze-out volume, we fix a 
volume and then calculate the caloric curve and thus the specific heat 
at  constant volume.
 
In the present calculation, we choose the system $^{150}$Sm which lies 
in the mass range of interest for the two experiments mentioned earlier. 
The freeze-out volume has been taken to be $V = 8 V_0$ where $V_0$ 
is the normal volume of the nucleus at zero temperature. 
This volume is close to the value generally used in statistical
multifragmentation models \cite{Gro,BBI} and to the one extracted from
the double ratio of the isotope yields with inclusion of radial flow
\cite{SDK}. The physical observables of interest like 
phase transition temperature is found to be very weakly dependent on the
freeze-out volume beyond $V = 8 V_0$ as noted earlier \cite{DDSS}. 
In Figure 1, we display the caloric curve for $^{150}$Sm. The 
excitation energy per particle $e^{*}$ is defined as 
$$
e^{*} = e(T)-e(T=0).
$$
The caloric
curves are shown for three values of pressures, namely, $P_0 =0,-0.05$ 
and -0.1 MeV $fm^{-3}$. The flow pressure$ P_f (= -P_0)$ is shown 
\cite{PSD} to be related to the flow energy as
$$
      P_f = D(v_f,T)\rho(r) e_f (r), 
\eqno (7)
$$
where the quantity $D(v_f,T)$ depends weakly on the temperature $T$ 
and the radially directed flow velocity $ v_f$ and is $\simeq$ 4.5 for 
neutrons or protons \cite{PSD} and $e_f(r)$ the flow energy per nucleon 
at any point within the volume. The total flow energy may be expressed as 
$$
E_f = \int_{V}\rho(r) e_f(r) d{\bf r}
    = P_f V/D(v_f,T). 
\eqno (8)
$$
 The average flow energy per nucleon is then $\simeq 1.3$ MeV for 
$P_0 = -0.1$ MeV $fm^{-3}$. When there is no flow ($P_0 = 0$), the caloric 
curve is smooth with initial faster rise of temperature with excitation 
energy, then a slower rise and lastly a kink at $T\simeq 10$ MeV, after 
which the excitation energy rises linearly with temperature. With 
increase in flow energy, the rise in temperature is slower and when the 
pressure $P_0 = -0.1$ MeV $fm^{-3}$, the caloric curve shows a plateau at 
$T\simeq 5$ MeV in the excitation energy range of 5-10 MeV. In the above
calculations, we have taken constant pressure independent of excitation
energy. One expects however increase in flow energy with increasing
excitation, which may influence a change in the caloric curve. To
investigate this aspect, we take the flow energy and hence the external
pressure proportional to $T^2$ as the excitation energy is not apriori
known. Guided by the experimental data \cite{Poch}  that the slope of the
linearly rising portion of the caloric curve is 3/2, we further assume that
the flow energy saturates at $T \simeq $ 5 MeV. A representative calculation
with the variable flow energy is also displayed in Figure 1 where the
flow pressure saturates at a value $P_0$ = -0.1 MeV $fm^{-3}$ at T=5.0
MeV. We find that the change in the caloric curve compared to that for
constant pressure of $P_0$ = -0.1 MeV $fm^{-3}$ is not significant.
In Figure 2, the corresponding specific heat defined as
$$
      C_v = (de^\ast/dT)_v,
\eqno (9)
$$
are displayed. The broad-based peak for no flow goes over to an extremely 
sharp peak from $T\simeq 9.5$ MeV to $T \simeq 5$ MeV with increasing 
 flow. The structure of $C_v$ and the other observables discussed
subsequently for the variable flow reaching a saturation as mentioned
are practically the same as those for constant pressure with the
saturation value and therefore are not displayed in our results. 
Looking at Figures 1 and 2, it appears that the system signals a 
liquid-gas phase transition at the peak temperatures and with increase 
in flow energy, the system moves from a continuous phase transition to 
a sharp first order phase transition. The linear rise of excitation 
energy with temperature beyond the phase transition temperatures for 
all the cases in Figure 1 reflects the fact that the system behaves 
like a noninteracting gas of classical particles; the value of the 
slope there is 3/2, the classical value for $C_v$. 
  
   In Figure 3, the rms radius of the proton distribution for the 
system is shown as a function of temperature. For all the three
cases with constant pressure, 
the radius initially increases very slowly with temperature
and then the rise is very fast  up to the phase transition temperature.
For the cases accompanied with flow energy, there is an
extremely sharp transition in the density distribution at the phase 
transition temperature as is evident from the very sharp increase in
the rms radius. We have earlier noted \cite{DDSS} that when there is
no flow, near the transition temperature within an interval of
$\Delta T\sim 0.5$ MeV, the density distribution loses its structure
and looks like a uniform distribution of matter inside the volume.
With inclusion of flow energy, within an interval of $\Delta T\sim 0.1$
MeV near the transition temperature, the density distribution now 
undergoes an exotic shape transition to a bubble shape at $T=5.3$ MeV
for $P_0 = -0.1$ MeV $fm^{-3}$ which is displayed in Figure 4. The 
appearance  of bubble-type configurations has also been noted earlier 
\cite{Wong,NBNT} in the dynamics of the expanding nuclei. A possible 
connection of these exotic shapes to first-order liquid-gas phase 
transition may be made from Figures 1 and 4.  The slow fall in rms 
radius beyond the transition temperature ( for non-zero flow ) is found 
to be due to the filling up of the interior of the bubble.
     
     We now make a comparison with the experimental data mentioned earlier. 
Both experiments refer to 
projectile fragmentation studies from far-central collisions, have little
flow energy $\approx 0-2$ MeV per nucleon \cite{LYN,DUR} and have a range 
of fragment masses studied from A $\approx 40$ to A $\approx$  200. 
We have therefore repeated the calculations
for a spectrum of masses in the mass range of experimental interest. We
find \cite{dss} nearly the same behaviour in the caloric curve with transition 
temperatures differing at most by $\approx  0.2$ MeV in the case with 
$P_0 = -0.1$
MeV $fm^{-3}$. The Au+Au data at 600 A.MeV fit extremely well with the
theoretical results with $P_0 = -0.1$ MeV $fm^{-3}$ which corresponds to
a flow energy of $\approx  1.3$ MeV per nucleon. The Au+C data at the 
higher energy of 1 A.GeV also fall in the band of theoretical results 
calculated with $P_0$ between $0.0$ to $-0.1$ MeV $fm^{-3}$. Here the 
target is very light
and a lesser compression or lower flow energy is expected. We calculated 
the incompressibility of the nuclei in the scaling model as a function
of temperature and found that in both cases with or without flow, the
incompressibility vanishes at the transition temperature which may
add further as a positive signal to liquid-gas phase transition \cite{dss}.
     
To summarise, we have calculated the caloric curve and the specific
heat of a finite nucleus with mass in the range of experimental 
interest in a self-consistent finite temperature Thomas-Fermi
theory with and without any collective radial flow and find signals
of liquid-gas phase transition in both cases. In the case without
flow, there appears to be a continuous phase transition at a 
temperature $T\simeq 9.5$ MeV, much below the critical temperature for 
infinite nuclear matter ($T_c \simeq 16$ MeV ) whereas with inclusion 
of a nominal flow energy of $\sim 1.3$ MeV per nucleon,
the transition appears to be a first order phase 
transition at  $T\simeq 5$ MeV. In the latter case, there is in 
addition a phase transition to a bubble shape from a diffuse spherical 
shape. The conclusions remain unaltered even if an excitation energy
dependent flow as mentioned earlier is employed. 
The bubble shape may be an artifact of the spherical symmetry
imposed in the FTTF calculation; if such constraints were relaxed, 
different shapes like a torus might have been obtained. The problem,
however, then becomes numerically more involved. The theoretical results
fit extremely well with the experimental data. 
In the higher energy 1 A GeV Au 
on the very light C nucleus, the projectile-like fragment suffers 
possibly very little or no compression and then the caloric curve 
alludes to a continuous phase transition.
For the lower energy 600 A MeV Au+Au,
slight compression leading to a modest radial flow energy may not
possibly be ignored and then  changes in the caloric curve may result 
as is indicated in the experiments.

\bigskip
 
The authors express their sincere thanks to Prof. S. Dasgupta 
discussions with whom initiated this work. This work is partially 
supported by the U.S. National Science Foundation under grant 
No. PHY-9413872.

\newpage

\begin{center}
{\bf FIGURE CAPTIONS}
\end{center}
\bigskip

{\bf Fig. 1}
The excitation energy per particle plotted as a function of temperature 
(caloric curve) for the system $^{150}$Sm at freeze-out volume $ V=8V_0$. 
The dotted, dashed and the full lines correspond to $P_0 =0.0, -0.05$
and -0.1 MeV $fm^{-3}$, respectively. The dash-dotted line corresponds 
to variable pressure (see text). The data are taken from Ref. [8]
(full squares) and from Ref. [15] (open circles). 
 
{\bf Fig. 2} 
The specific heat per particle plotted as a function of temperature for
the system $^{150}$Sm with $ V=8V_0$. The dotted, dashed and the full 
lines have the same meaning as in Fig. 1.

{\bf Fig. 3} 
The proton rms radius is plotted as a function of temperature for the
system $^{150}$Sm. The symbols have the same meaning as in Fig. 1.

{\bf Fig. 4}
The proton density profile for the system $^{150}$Sm calculated with
$ P_0 = -0.1$ MeV $fm^{-3}$ and $V=8V_0$. The dashed and the full lines 
correspond to $T$= 5.2 and 5.3 MeV, respectively. 

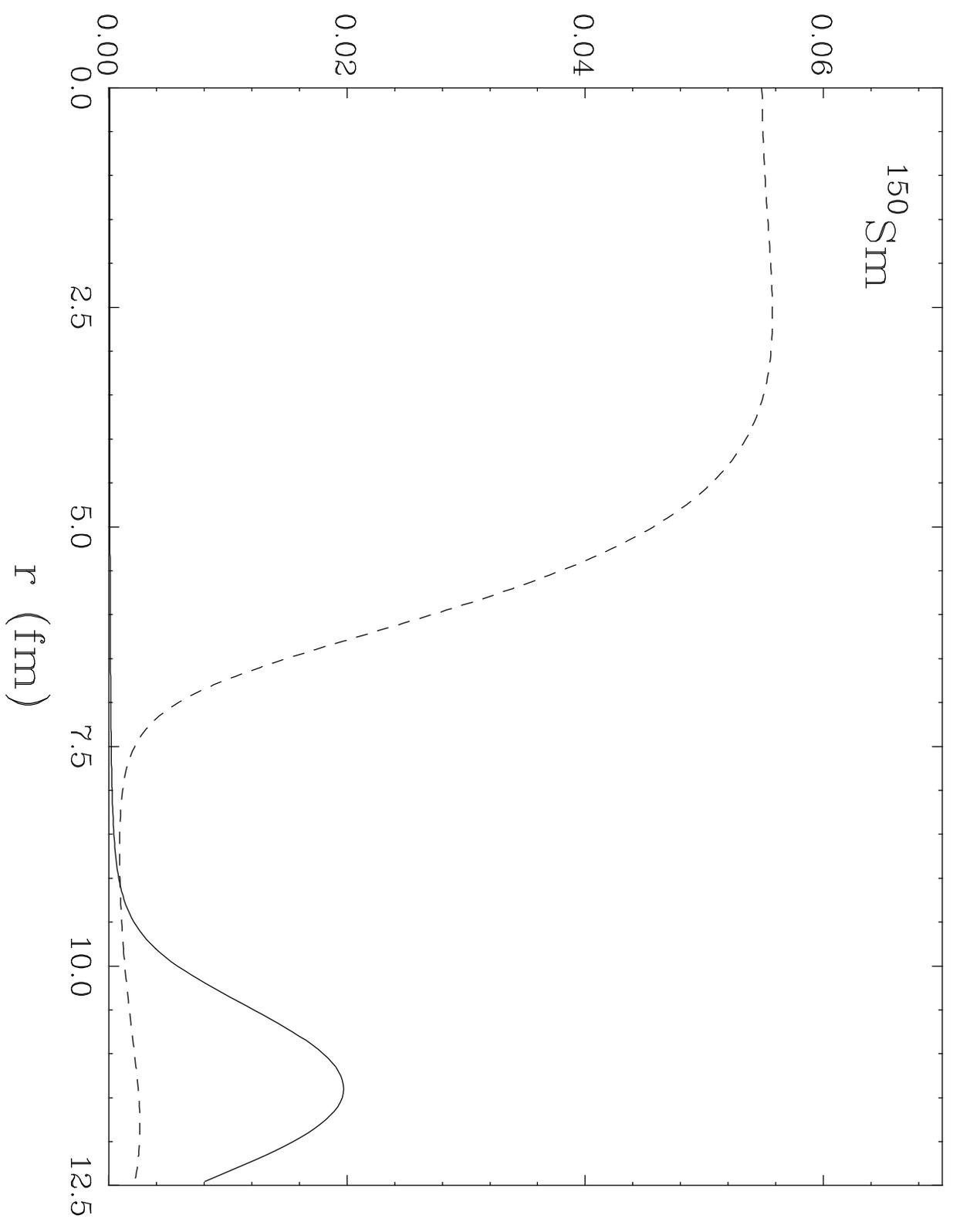
\end{document}